\documentclass[paper]{JHEP3}

\usepackage{latexsym}
\usepackage{amsmath}
\usepackage{amscd}
\usepackage{amsfonts}
\usepackage{amsthm}
\usepackage{epsfig}
\allowdisplaybreaks[1]

\def\IN{\mathbb{N}}

\newcommand{\Tr}{{\rm Tr}}
\def\ap#1{\alpha^{\prime\,#1}}

\title{The non-abelian D-brane effective action
through order $\alpha'{}^4$}
\author{Paul Koerber\thanks{Aspirant FWO} and Alexander
Sevrin\\
Theoretische Natuurkunde, Vrije Universiteit Brussel \\
Pleinlaan 2, B-1050 Brussels, Belgium\\
E-mail: \email{koerber@tena4.vub.ac.be}, \email{asevrin@tena4.vub.ac.be}}
\preprint{VUB/TENA/02/05\\ \hepth{0208044}}

\abstract{Requiring the existence of certain BPS solutions to the equations
of motion, we determine the bosonic part of the non-abelian D-brane
effective action through order $\alpha'{}^4$. We also propose an economic organizational
principle for the effective action.}

\keywords{D-branes}

\begin{document}
\section{Introduction}
The bosonic worldvolume
degrees of freedom of a single D$p$-brane are $9-p$ scalar fields and a
$U(1)$ gauge field in $p+1$ dimensions \cite{pol}. In the limit of
slowly varying fields, the effective action which determines the dynamics of the D-brane at low energies,
is known to all orders in $\alpha '$. It is the ten-dimensional
supersymmetric Born-Infeld action, dimensionally reduced to $p+1$
dimensions \cite{BI}. Its supersymmetric extension was obtained in
\cite{susynbi}. The knowledge of the full effective action was crucial
for numerous applications.

Once several, say $n$, D-branes coincide, the gauge group is enhanced
from $U(1)$ to $U(n)$, \cite{witten}. The full non-abelian extension of the
Born-Infeld theory is not known yet. This is mainly due to two (related) reasons. As all
fields take their values in the adjoint representation of $U(n)$, an ordering
prescription is needed. Besides this we also need to include derivative terms.
Indeed the limit of (covariantly) constant fields is intrinsically related to
the abelian limit as $D_c F_{ab}=0$ implies that $[F_{dc},F_{ab}]=0$.
This can be seen in another way. In the abelian
case, the effective action consists of terms of the form (in a very schematic notation)
$g^{-2}\alpha'{}^{2m-2} \partial^{2n} F^{2m-n}$. Performing the following rescaling\footnote{The parameter
$\beta$ reflects the only scaling freedom left. It can e.g. conveniently be chosen such that $g$ and
$\alpha'$ remain fixed in the limit $\varepsilon\rightarrow 0$.},
$x\rightarrow \varepsilon^{-1} \beta^{-1}x$, $A\rightarrow\beta A$, $\alpha'\rightarrow
\varepsilon^{-1}\beta^{-2}\alpha'$, $g\rightarrow \varepsilon\beta^2g$, we find that
$g^{-2}\alpha'{}^{2m-2} \partial^{2n} F^{2m-n}\rightarrow
\varepsilon^n g^{-2}\alpha'{}^{2m-2} \partial^{2n} F^{2m-n}$. Sending $\varepsilon\rightarrow 0$,
the derivative terms vanish and the Born-Infeld action remains.
In the non-abelian case we still have that $g^{-2}\alpha'{}^{2m-2} D^{2n} F^{2m-n}\rightarrow
\varepsilon^n g^{-2}\alpha'{}^{2m-2} D^{2n} F^{2m-n}$. However we also have that
$D\cdot= \partial\cdot+[A,\cdot]\rightarrow\varepsilon\beta(\partial\cdot+\varepsilon^{-1}[A,\cdot])$ and
$F= \partial A+[A,A]\rightarrow \varepsilon\beta^2(\partial A+\varepsilon^{-1}[A,A])$
which makes the $\varepsilon\rightarrow 0$ limit meaningless. Other arguments for the relevance of the derivative
terms in both the abelian and non-abelian case were given in \cite{bilal1}.

At this moment the non-abelian effective action is known through $ {\cal O}(\alpha'{}^3)$ including
the terms quadratic in the gauginos. The leading order term of the effective action for $n$
coinciding D$p$-branes is the ten-dimensional $N=1$ supersymmetric
$U(n)$ Yang-Mills theory dimensionally reduced to $p+1$ dimensions. There are no ${\cal O}(\alpha')$ corrections.
The bosonic ${\cal O}(\alpha'{}^2)$ were first obtained in \cite{direct} and \cite{direct1}
while the fermionic terms were obtained in \cite{goteborg} and \cite{bilal}.
In \cite{goteborg} supersymmetry fixed the correction while in \cite{bilal} a direct calculation starting from
four-point open superstring amplitudes was used. At higher orders a direct calculation becomes problematic\footnote{
We thank Stefan Stieberger for explaining this to us.} and one has to rely on indirect methods. A very powerful strategy
was proposed in \cite{lies}\footnote{A very different approach was recently proposed in \cite{howe}.}.

Central in the approach of \cite{lies}, was the existence of higher-dimensional generalizations of instantons:
stable holomorphic bundles. They solve the Yang-Mills equations of motion and in a D-brane context they correspond
to BPS configurations. The effective action can be viewed as a deformation of the Yang-Mills action. Requiring that
stable holomorphic bundles, or some deformation thereof, solve the equations of motion yields a recursive method to
construct the effective action. While this approach becomes very tedious at higher orders, it has the great advantage
that its algorithmic nature allows for a computerized approach. To this end a program in Java, an object oriented
language based on the syntax of C, was developed \cite{paul}.
In \cite{sk1} the method was successfully applied to determine
the bosonic ${\cal O}(\alpha'{}^3)$ terms in the effective action.
Very recently, in \cite{groningen}, supersymmetry was used not only to confirm the results of \cite{sk1} but to
construct the terms quadratic in the gauginos through this order as well. The full effective action through this
order was tested in \cite{test} and there is no doubt left that the result is indeed correct.

In the present paper we extend the results of \cite{sk1} and we obtain the bosonic terms in the effective action at
order $ \alpha'{}^4$. Furthermore we will propose a very economic way to organize the terms in the effective action.

Throughout the paper we will put $2\pi\alpha'=1$.

\section{General strategy}
\label{strategy}
We consider a $U(n)$ Yang-Mills theory in $2p$ dimensions with a Euclidean signature. Its equations
of motion are given by\footnote{Throughout this paper, we write all indices down and
we sum over repeated (real and complex) indices.},
\begin{eqnarray}
\label{eom0}
D_a F_{ab}=0.
\end{eqnarray}
Switching to complex coordinates $z^\alpha=(x^{2\alpha-1}+ix^{2\alpha})/\sqrt{2}$,
$\bar z^{\bar\alpha}=(x^{2\alpha-1}-ix^{2\alpha})/\sqrt{2}$, eq.\ (\ref{eom0}) reads,
\begin{eqnarray}
0&=&D_{\bar\alpha }F_{\alpha \bar\beta}+D_{\alpha }F_{\bar\alpha
\bar\beta}\nonumber\\
&=&D_{\bar\beta}F_{\alpha \bar\alpha }+2D_{\alpha }F_{\bar\alpha
\bar\beta},
\end{eqnarray}
where we used the Bianchi identities. One sees that imposing the following linear
relations between the field-strengths,
\begin{eqnarray}
\label{hol}
F_{\alpha\beta}=F_{\bar\alpha\bar\beta}=0,
\end{eqnarray}
and
\begin{eqnarray}
\label{duy}
\sum_\alpha F_{\alpha\bar\alpha}\equiv F_{\alpha\bar\alpha}=0,
\end{eqnarray}
provides a solution to the equations of motion. In four dimensions, these are exactly
the standard instanton equations. In general, eq.\ (\ref{hol}) defines a holomorphic bundle
and eq.\ (\ref{duy}) requires it to be stable \cite{duy}. These solutions
were discovered in \cite{mons}.

Following \cite{sk1}, we subsequently construct, order by order in $\alpha{}'$, all possible terms of the
correct dimension which can be build out the field-strength and its derivatives. More concretely, at order $\alpha'{}^m$,
we write down all possible terms having $2n$ covariant derivatives, with $n\in\{0,1,\cdots m\}$, and $m-n+2$
field-strength tensors, taking the cyclicity of the group trace into account. Each of these terms gets an arbitrary
coupling constant. The same game has to be played for the most general deformation of eq.\ (\ref{duy}), which
through this order has the form
\begin{eqnarray}
\label{duy1}
F_{\alpha\bar\alpha}=\sum_{n=1}^{m} {\cal F}_{(n)},
\end{eqnarray}
where $ {\cal F}_{(n)}$ with $n<m$ were already determined at lower orders and $ {\cal F}_{(m)}$ is the most general polynomial
of terms having $2n$ derivatives and $m-n+1$ field-strengths, where $n\in\{0,1\cdots,m\}$.
Here again we leave the coupling constants free. Surprisingly, the coupling constants in {\em both} the lagrangian and the stability
condition are determined by requiring that eqs.\ (\ref{hol}) and (\ref{duy1})
solve the equations of motion. The presence of derivative terms however, complicates
the analysis considerably. Indeed when writing down the most general lagrangian at a certain order in $\alpha'$, the following
points have to be taken into account,
\begin{enumerate}
\item Terms can be related through partial integration.
\item Terms can be related by Bianchi identities.
\item The $[D,D]\cdot=[F,\cdot]$ identities\footnote{In their most general form they
read: $[D_1D_2\ldots D_{n-2}F_{n-1,n},D_{n+1}\ldots D_{n+m-2}F_{n+m-1,n+m}]  = $
$[D_1,[D_2,\ldots [D_{n-2},[D_{n-1},D_n]]\ldots]]D_{n+1} D_{n+m-2}F_{n+m-1,n+m}$.
%\beq
%\begin{split}
%[D_1D_2\ldots D_{n-2}F_{n-1,n},D_{n+1}\ldots D_{n+m-2}F_{n+m-1,n+m}] & = \\
%[D_1,[D_2,\ldots [D_{n-2},[D_{n-1},D_n]]\ldots]]D_{n+1} D_{n+m-2}F_{n+m-1,n+m} & .
%\end{split}
%\eeq
But we will use the above shorthand in the rest of the paper.}
relate certain terms.
\item Certain terms can be eliminated by field redefinitions.
\end{enumerate}
Similar considerations hold, apart from the first point, for the stability condition.
We discard the fourth point until we perform the final analysis of the
resulting action. The reason for this is that certain field redefinitions affect the complex structure
such that eq.\ (\ref{hol}) gets modified and, as a consequence, our method is not entirely field redefinition
independent.

Because of the identities of points one, two and three, many terms are dependent.
Handling this essentially boils down to choosing an appropriate basis of independent terms
in the lagrangian as well as in the stability condition.  The basis terms in the lagrangian must be
such that there is a neat separation between the irrelevant field redefinition terms and the
relevant non field redefinition terms.  In concreto, when expressing the result in a certain basis, the coefficients
of some of the basis terms can be shifted by field redefinitions where, again, great care is needed because
the field redefinitions are also related by the identities of point two and three.
The shifts are independent when the number of such field redefinition terms is minimal.
In that case, we can bring their coefficients to zero.
So only the projection of the result on the non field redefinition terms is relevant.

Of course, there is still a lot of freedom in choosing a basis, so we need some kind of organizational
principle to get rid of at least part of this freedom in an economic way.

\section{Organizational principle}
\label{organization}

As in the abelian case, it is possible to classify terms at a certain order in $\alpha'$ according to
the number of derivatives.  For this, we must get rid of the $[D,D]\cdot=[F,\cdot]$ identities.
Consider any linear combination of terms.
Now, start with the terms without derivatives and fully symmetrize in the fieldstrength tensors.
In this process, we use the $[D,D]\cdot=[F,\cdot]$ identities to convert the introduced commutators
of fieldstrengths to commutators of derivatives\footnote{This means that we push the $[D,D]\cdot=[F,\cdot]$ identities
to the right.
The reader could wonder why we do not push those identities to the left
and use for instance symmetrized derivatives.  This leads indeed to fewer terms and fewer identities,
because there are no antisymmetrized Bianchi identities.
On the other hand the remaining partial integration and Bianchi identities are more complicated
and the final result is awful because we get no clear separation between field redefinition and non
field redefinition terms}.
Next, we turn to the terms with two derivatives an again fully symmetrize
them, whereby a term of the form $DF$ or $D^2F$ is considered as a single entity.
Again, terms with more derivatives are added as compensation.
We proceed in this fashion order by order in the number of derivatives.
Since the resulting terms are symmetrized in the fieldstrength tensors, all ``non-abelianality'' sits in the
covariant derivatives.  From now on all terms, whether they be part of the lagrangian, stability condition, field
redefinitions or equations of motion, should be thought of as symmetrized. Note that this approach follows the spirit
of \cite{arkady} and we could call it a ``generalized symmetrized trace prescription''.

A major advantage is that the non-abelian calculation follows the abelian one more closely, since
symmetrized terms in the lagrangian lead to symmetrized terms in the equations of motion.
Except for the obvious fact that the derivatives are non-commuting there are however some
other differences:
\begin{enumerate}
\item There are new identities, because in symmetrizing we only used up part of the
${[}D,D{]} \cdot = {[}F,\cdot{]}$ identities.  An example will clarify this.
Only one of the identities\protect\footnote{When writing down long equations or results in the rest of the paper, 
we will use a shorthand notation for the indices i.e.\
$1,2,3,\ldots$ instead of $a_1,a_2,a_3,\ldots$ Repeated indices are still summed over.}
\begin{eqnarray}
{[}F_{12}, F_{34}{]} & = & {[}D_1, D_2{]} F_{34} \nonumber\\
{[}F_{34}, F_{12}{]} & = & {[}D_3, D_4{]} F_{12} \, ,
\end{eqnarray}
is used to commute $F_{12}$ and $F_{34}$.  The other one is left in the form:
\begin{eqnarray}
{[}D_1, D_2{]} F_{34} + {[}D_3, D_4{]} F_{12} = 0 \, .
\end{eqnarray}
These kind of identities are related to antisymmetry, 
as in this case, or to Jacobi identities of fieldstrength commutators.
In general they read:
\begin{eqnarray}
&& {[}D_1, {[}D_2, \ldots ,{[}D_{m-2}, {[}D_{m-1},D_{m}{]}{]}\ldots{]}{]} D_{m+1} D_{m+2} 
\ldots D_{m+n-2} F_{m+n-1,m+n} + \nonumber\\
&& {[}D_{m+1}, {[}D_{m+2}, \ldots ,{[}D_{m+n-2}, {[}D_{m+n-1},D_{m+n}{]}{]}\ldots {]}{]} 
D_1 D_2 \ldots D_{m-2} F_{m-1,m} = 0 \, , 
\nonumber\\
\end{eqnarray}
and we call them {\em antisymmetrized Bianchi identities}.
\item Consider the stability condition\protect\footnote{The order $\alpha'$ correction is zero.} eq. (\ref{duy1}):
\begin{eqnarray}
F_{\alpha\bar{\alpha}} - {\cal F}_{(2)} - {\cal F}_{(3)} - \cdots = 0 \, .
\end{eqnarray}
Somewhere in the equations of motions we will find $\mbox{Sym} \{F_{\alpha\bar{\alpha}} T\}$, where $\mbox{Sym}$
means ``symmetrized in the fieldstrength tensors'' and $T$ contains, say, $n$ fieldstrengths.
For this to be zero, and thus for the stability condition to solve this piece of the equations of motion, the term
$-\mbox{Sym} \{({\cal F}_{(2)} )T\}$ must also be present in the equations of motion at higher order.
Here the inner brackets mean that the factors of ${\cal F}_{(2)}$ stay together, so we must
further symmetrize this term by mixing those factors among the factors of $T$.  Unlike in the
abelian case, terms with more derivatives are thus introduced.  Interestingly, the number of extra 
derivatives in these terms must be a multiple
of four. Indeed, both $\mbox{Sym}\{({\cal F}_{(2)}) T\}$ and $\mbox{Sym} \{ {\cal F}_{(2)} T \}$
are symmetric under the reversion of all factors, so their difference 
is symmetric as well and hence must contain an {\em even} number of commutators.

This is the only way in which terms without derivatives --- and using also only contributions without
derivatives to the stability condition --- communicate with terms with derivatives and thus
the mechanism that prevents the ``ordinary'' symmetrized trace to have BPS solutions.
\end{enumerate}

After we found the result in this way, we used a second organizing principle to simplify even further.
We tried to use basis terms with as many ``groups'' of nested covariant derivative commutators
as possible.  Each group corresponds, using a $[D,D]\cdot=[F,\cdot]$ identity to a commutator of fieldstrengths or equivalently to
an algebra structure constant which can be put in front.
To clarify this correspondence we write result \eqref{l3} in different ways using $[D,D]\cdot=[F,\cdot]$ identities:
\begin{equation}
\begin{split}
 [D_3,D_2] D_4 F_{51} D_5 [D_4,D_3] F_{12}  & = \\
 [F_{32},D_4 F_{51}]D_5 [F_{43},F_{12}] & = \\
 [D_4,[D_5,D_1]]F_{32} D_5 [D_1,D_2]F_{43} & \, .
\end{split}
\end{equation}
The first and the third line show that the number of {\em groups} of commutators of 
derivatives is important rather than the number of commutators.
Although we pushed the $[D,D]\cdot=[F,\cdot]$ identities to the right initially,
in this way we can easily see which terms can be pushed how far in the other direction if necessary.
This might be useful when comparing to results following from string amplitude calculations, as
they tend to give results where commutators of derivatives are pushed into antisymmetric combinations
of fieldstrength tensors.
\DOUBLEFIGURE{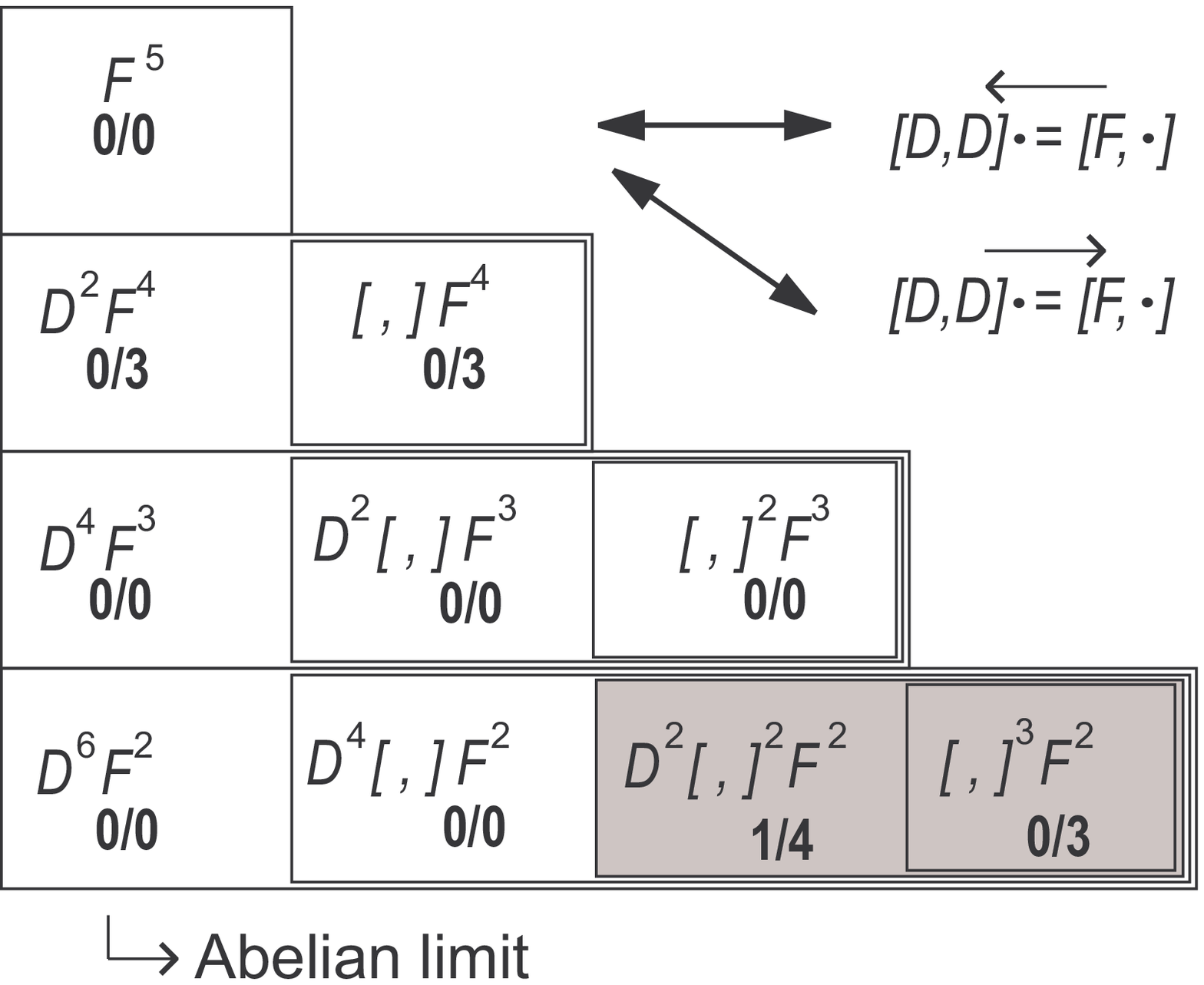,scale=0.37}
{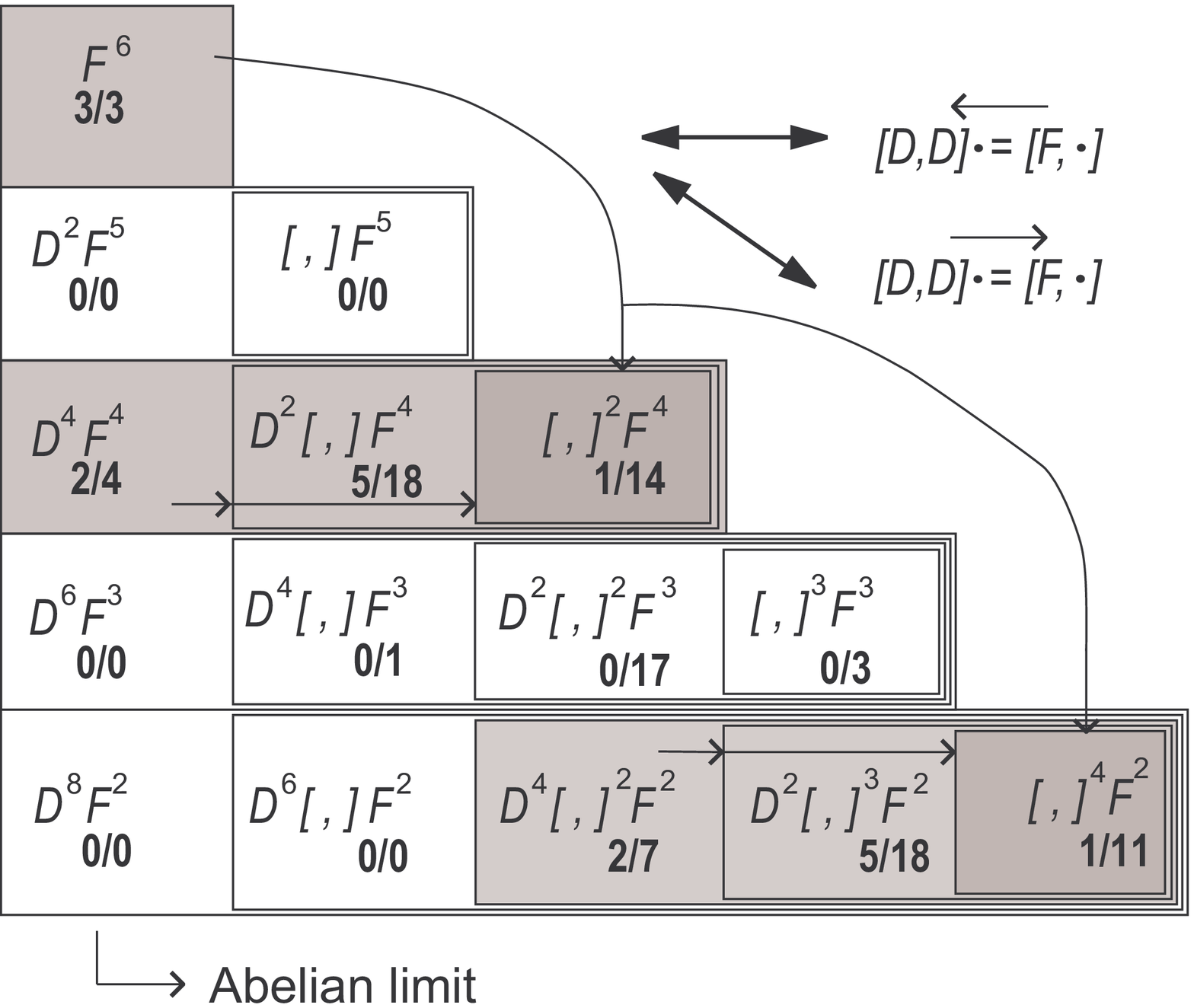,scale=0.37}
{\label{alpha3}Basis terms at order $\alpha'{}^3$. The horizontal classification is by the number of derivatives if the $[D,D]\cdot=[F,\cdot]$ identities are
pushed to the left as in the text. The vertical classification is by the number of commutator groups of $D$s.  Note
that when pushing the $[D,D]\cdot=[F,\cdot]$ identities to the right, you can find the terms with the same number
of derivatives on the diagonals as indicated. The numbers indicate (\#non-zero coefficients in result)/(\#all non field redefinition basis terms). The
grey area indicates the basis terms potentially used for the result.  We tried to simplify as much as possible
and put as many as possible coefficients to zero.}
{\label{alpha4}Basis terms at order $\alpha'{}^4$. See also the caption of figure \ref{alpha3}. The arrows indicate how the
abelian result transforms in the non-abelian result. Partial integration and Bianchi identities mix terms with different
numbers of commutator groups.  Hence the vertical arrows.  For the diagonal arrows, see the argument 
of section \ref{organization}, point 2.
Using our method in the abelian case, there are two groups of terms, the Born-Infeld part and the 
$\partial^4 F^4$ part.  The
latter has an arbitrary constant.  Note also the new group of terms with 8 derivatives. The dark grey areas are where different groups meet
in the full non-abelian case.  As a result the arbitrary constant is fixed.}

See figure \ref{alpha3} and \ref{alpha4} for the classification of the terms.
Obviously, there is still basis freedom left, so that some terms in the result can move to the right.
E.g. result \eqref{l3} can also be written as a sum of terms with $2$ and $3$ structure constants\footnote{See the result in \cite{groningen}.}, however
not with terms with only $3$ structure constants. Hence the grey area in figure \ref{alpha3}.

\section{The result}
\label{result}
The purely bosonic part of the non-abelian effective action through $ {\cal O}(\alpha'{}^4)$ is of the form,
\begin{eqnarray}
{\cal L}=\frac{1}{g^2}\left({\cal L}_0+{\cal L}_2+{\cal L}_3+{\cal L}_4\right),
\end{eqnarray}
where the leading term is simply\footnote{We use the following notation:
$F^m\equiv F_{a_1a_2}F_{a_2a_3}\cdots F_{a_ma_1}\equiv F_{12}F_{23}\cdots F_{m1}$.}
\begin{eqnarray}
\label{l0}
{\cal L}_0=\;=\; -
 \Tr\, \left\{\frac{1}{4}F^2\right\}\,.
\end{eqnarray}
Subsequently we have
\begin{eqnarray}
\label{l2}
{\cal L}_2=\mbox{STr} \left\{\frac{1}{8} F^4
- \frac{1}{32} F^2F^2
\right\},
\end{eqnarray}
where STr denotes the symmetrized trace prescription. At this point both the overall multiplicative factor in front of the
action as well as the scale of the gauge fields got fixed \cite{sk1}. The next term is\footnote{All results are of course modulo
field redefinition terms.}
\begin{eqnarray}
\label{l3}
{\cal L}_3=\frac{\zeta (3)}{2\pi^3}\Tr\left\{[D_3,D_2] D_4 F_{51} \, D_5 [D_4,D_3] F_{12} \right\}.
\end{eqnarray}
In fact our method did not fix the overall multiplicative factor in front of this term. As explained in
\cite{sk1} and \cite{groningen}, this was most fortunate as one expects from string theory that the
coupling constants of terms which are of odd order in $\alpha'$ are irrational while our method yields,
at this order, only rational coupling constants. We determined the coupling constants by comparing the relevant terms
to the partial results in \cite{bilal1}.
Note that eq.\ (\ref{l3}) looks very different from the expression given in \cite{sk1}. The reason for this is clear.
In \cite{sk1}, we used precisely the opposite organizational strategy as the one proposed in section \ref{organization}.
Comparing eq.\ (\ref{l3}) to the corresponding expression in \cite{sk1} (or \cite{groningen} where yet another basis was
chosen) shows convincingly that the scheme proposed in section \ref{organization} is the most economical.

We now turn to the new result of this paper. The fourth order contribution to the effective action reads (see figure \ref{alpha4})
\begin{eqnarray}
{\cal L}_4 = {\cal L}_{4,0} + {\cal L}_{4,2} + {\cal L}_{4,4} \, ,
\end{eqnarray}
with
\begin{equation}
\begin{split}
{\cal L}_{4,0} & =  \mbox{STr} \left( \frac{1}{12} F_{12}F_{23}F_{34}F_{45}F_{56}F_{61}
                 - \frac{1}{32} F_{12}F_{23}F_{34}F_{41}F_{56}F_{65}
                  + \frac{1}{384} F_{12}F_{21}F_{34}F_{43}F_{56}F_{65} \right) , \\
{\cal L}_{4,2} & = \frac{1}{48} \mbox{STr} \Big( -2 \, F_{12}D_{1}D_{6}D_{5}F_{23}D_{6}F_{34}F_{45}
                                     - F_{12}D_{5}D_{6}F_{23}D_{6}D_{1}F_{34}F_{45}  \\
               & + 2 \, F_{12}\left[D_{6},D_{1}\right] D_{5}F_{23}F_{34}D_{4}F_{56}
                 + 3 \, D_{4}D_{5}F_{12}F_{23}\left[D_{6},D_{1}\right]F_{34}F_{56} \\
               & + 2 \, D_{6}\left[D_{4},D_{5}\right]F_{12}F_{23}D_{1}F_{34}F_{56}
                 + 2 \, D_{6}D_{5}F_{12}\left[D_{6},D_{1}\right]F_{23}F_{34}F_{45} \\
               & + 2 \, \left[D_{6},D_{1}\right]D_{3}D_{4}F_{12}F_{23}F_{45}F_{56} \\
               & + \left[D_{6},D_{4}\right]F_{12}F_{23}\left[D_{3},D_{1}\right]F_{45}F_{56} \Big) , \\
{\cal L}_{4,4} & = \frac{1}{1440} \mbox{STr} \Big( D_6 [D_4,D_2]D_5 D_5 [D_1,D_3] D_6 F_{12} F_{34}
                   + 4 \, D_2 D_6 [D_4,D_1][D_5,[D_6,D_3]] D_5 F_{12} F_{34}  \\
                 & + 2 \, D_2 [D_6,D_4][D_6,D_1] D_5 [D_5,D_3] F_{12} F_{34}
                   + 6 \, D_2 [D_6,D_4]D_5[D_6,D_1][D_5,D_3] F_{12} F_{34} \\
                 & + 4 \, D_6 D_5 [D_6,D_4][D_5,D_1][D_4,D_3] F_{12} F_{23}
                   + 4 \, D_6 D_5 [D_4,D_2][D_6,D_1][D_5,D_3] F_{12} F_{34} \\
                 & + 4 \, D_6 [D_5,D_4][D_3,D_2][D_5,[D_6,D_1]] F_{12} F_{34} \\
                 & + 2 \, [D_6,D_1][D_2,D_6][D_5,D_4][D_5,D_3]F_{12}F_{34} \Big) \, .
\end{split}
\end{equation}

The terms with zero derivatives, ${\cal L}_{4,0}$, form of course the symmetrized trace Born-Infeld action, while the abelian limit
of the terms with four derivatives, the first two terms of ${\cal L}_{4,2}$, can be shown to agree with \cite{abelian4derivative}
\footnote{Although it is a little involved, this can still be shown by hand.  See e.g.\ Appendix B of \cite{wyllard}.}.  If we use our method in the abelian case at
order $\alpha'{}^4$, these terms have an arbitrary constant, since terms with a different number of derivatives have
no way of communicating in this case.  In the non-abelian case however, the coefficient is fixed at precisely the right value !

Note that there are no terms with two derivatives nor with six derivatives.  If there were, they would have had an arbitrary
constant since, following the reasoning of point 2 in section \ref{organization}, the symmetrized trace does not
communicate directly with terms in which the number of derivatives is not a multiple of four.

\section{Conclusions}
In the present paper we calculated the bosonic $\alpha'{}^4$ contribution to the non-abelian effective
D-brane action. As was already obvious from the $ {\cal O}(\alpha'{}^3)$ calculation, \cite{sk1}, \cite{groningen},
the inclusion of derivative terms is unavoidable. The ``generalized symmetrized trace prescription''
proposed in section \ref{organization} gives rise to a very economical way of organizing the action
which has the additional advantage that it closely mimics the abelian case.

When determining the coupling constants in the effective action and the deformed stability condition, we had to
solve 1816 equations for 546 unknowns. The fact that we found a unique (modulo field redefinitions) solution
consists in itself a strong check on our calculation. An independent check would follow the lines proposed in
\cite{HT}, where the mass spectrum in the presence of constant magnetic background fields was calculated from
the effective action and compared to the string theoretic result. A careful analysis in \cite{DST}
showed that the symmetrized trace prescription without derivative terms deviated from the string theoretical
results, starting at order $ \alpha'{}^4$. The derivative corrections which were obtained in this paper should cure
this problem. We postpone this check to a future paper. In this context we should also mention the analysis in
\cite{STT} where the mass spectrum in the presence of constant magnetic background fields combined with the requirement
that the abelian limit was correctly reproduced, were used to determine the effective action as far as possible.
Besides the fact that this method yielded a multi-parameter solution an important ansatz was made: only terms without
derivatives were taken into account. Our present paper clearly indicates that this ansatz was too strong.

The present result opens the way to all-order statements. E.g., we already noticed, within our generalized
symmetrized trace prescription, that terms without derivatives only communicate with terms with 4, 8, ... derivatives.
It follows that terms with 2, 6, 10, ... derivatives have arbitrary constants in our method.
It might very well turn out that the constants for the terms with 2 derivatives vanish and that these terms are zero,
just as in the abelian case \cite{abelian4derivative}.
As a consequence, terms with $2+4n$, with $n\in\IN$, could vanish as well for the even orders of $\alpha'$. The all-order lessons we can draw from our
method is presently under investigation and we will come back to this in a separate publication. Related to this,
a careful analysis of the deformed stability condition has been started.

Finally, we hope that our paper can shed some light on the remarkable results recently obtained in \cite{stieberger}.

\bigskip

\acknowledgments

\bigskip

The authors are supported in part by the ``FWO-Vlaanderen'' through project
G.0034.02, in part by the Federal Office for Scientific, Technical
and Cultural Affairs through the Interuniversity Attraction Pole P5/27 and
in part by the European
Commission RTN programme HPRN-CT-2000-00131, in which the authors are
associated to the University of Leuven.

\end{document}